# Digital unzipping of DNA through a solid-state nanopore: A theoretical study for base-by-base ratcheting

Tomoki Ohkubo*
Technology Research Laboratory, Shimadzu Corporation, 3-9-4 Hikaridai, Seika-Cho, Soraku-Gun, Kyoto, 619-0237, Japan
*Correspondence: tohkubo@shimadzu.co.jp

## Abstract

Solid-state nanopore DNA sequencers present mechanical and chemical stability, reusability, and large-scale integrability. However, their development is hindered by the absence of a protein-free mechanism for controlling DNA translocation, which is accomplished by motor proteins in their biological counterparts. Here, we propose and theoretically analyze a protein-independent ratchet mechanism based on the unzipping of double-stranded DNA at the nanopore rim. When the transmembrane bias exceeds a certain threshold, the base pairs mechanically dissociate, allowing one strand to translocate while the other remains upstream. This unzipping process is known to slow DNA motion, suggesting that voltage pulses can trigger individual unzipping events at externally defined times, a concept referred to as *digital unzipping*. However, the intrinsic unzipping barrier is insufficient to provide the dwell times required for a reliable ionic-current readout; therefore, an additional mechanism is needed to hold the DNA in place between voltage pulses. To overcome this limitation, we introduce a reversible hold mechanism implemented via electrostatic attraction between DNA and a charged nanopore wall, which temporarily immobilizes the strand once the unzipping fork catches on the nanopore rim. Using a statistical-mechanical model, we track the evolution of the mean and variance of DNA position through each ratchet cycle. Analytical expressions for the corresponding error probabilities show that submicrosecond switching of the hold mechanism enables base-by-base stepping with an error rate <5%. These results theoretically demonstrate that digital unzipping combined with a reversible hold mechanism can yield deterministic single-base motion, thus opening a viable route toward all-solid-state nanopore sequencing.

## Main

Solid-state nanopore sequencers can offer major advantages over biological nanopore sequencers,[1] including mechanical and chemical stability, reusability, long shelf-life, and compatibility with semiconductor microfabrication.[2–4] They are also free from the read-length limitations that arise for their biological counterparts when a motor protein dissociates from DNA.[1]

Despite these potential benefits, however, fully functional solid-state nanopore sequencers have not yet been demonstrated. The principal obstacle is the absence of a *ratchet* mechanism that moves DNA through a nanopore one nucleotide at a time, enabling accurate basecalling from noisy ionic-current signals—the role fulfilled by motor proteins in biological nanopore sequencers.

At present, the only ratcheting mechanism realized is sequence-by-expansion (SBX) technology, which enzymatically converts DNA into a polymer (Xpandomer) carrying bulky side groups that sequentially catch on the nanopore rim.[5] Although SBX achieves ratcheting without motor proteins, Xpandomer synthesis still depends on engineered DNA polymerases, limiting the read length to only a few kilobases. In addition, epigenetic information such as DNA methylation is lost during this process.

To eliminate protein dependence, we propose a ratchet mechanism based on the unzipping of double-stranded DNA (dsDNA). When the overhanging DNA is pulled into a nanopore narrower than 2 nm, the duplex region encounters an unzipping barrier at the pore rim.[6–10] Once the transmembrane bias exceeds a certain threshold, forced unzipping occurs, allowing one strand to translocate, whereas the other remains on the *cis* side. This mechanical unzipping process is known to involve a significantly slow translocation. These characteristics suggest that applying a transmembrane voltage in discrete pulses could, in principle, trigger the dissociation of individual base pairs at controlled times,[11] a concept referred to as *digital unzipping*.

Although conceptually simple, digital unzipping has not yet been experimentally realized, mainly because the small intrinsic unzipping barrier[12] leads to short residence times (tens of nanoseconds) that are far below those required for accurate basecalling.[1] Increasing the friction between DNA and the nanopore will extend the residence time, but achieving the required $10^4$–$10^5$-fold increase is far beyond what is currently feasible.

Here, we propose a protein-independent ratchet mechanism for solid-state nanopore sequencing in which a reversible *hold* mechanism, that is, an electrostatic attraction that can temporarily immobilize DNA against the pore wall, compensates for the intrinsic unzipping barrier of dsDNA, enabling base-by-base translocation without motor proteins. Figure 1(a) illustrates the conceptual design. The conductive layer embedded within the nanopore membrane is positively charged to attract and temporarily immobilize DNA, thereby functioning as the hold mechanism. Similar concepts were previously examined in simulations by Luan et al.[13] and Shankla and Aksimentiev,[14] who demonstrated the submicrosecond reversible trapping and release of DNA. Our focus here is not the engineering implementation but demonstrating, through theoretical analysis, that such a mechanism enables digital unzipping, that is, the base-by-base translocation of DNA.

The unzipping potential landscape is schematically shown in Fig. 1(b). At room temperature, the duplex form of DNA is more stable than the single strand; hence, the potential increases toward the *trans* side, favoring rezipping.[12] Applying a transmembrane bias tilts the potential to the right; at a certain *stall voltage* $V_{\text{stall}}$, the electrophoretic force balances the rezipping tendency and electrophoretic force, rendering the mean DNA position stationary. Increasing the transmembrane voltage beyond the stall voltage eventually eliminates the potential barrier at a *critical voltage* $V_c$, and the DNA begins to unzip and translocate continuously in the *drift* regime.

Figure 2 illustrates the timing sequence of the transmembrane voltage and hold mechanism. Both can be toggled on and off, resulting in four phases: the first hold, drift, second hold, and trap phases [Fig. 2(a)]. In the first hold phase, the positively charged conductive layer attracts DNA to the pore wall, immobilizing it, and the probability distribution freezes [Fig. 2(b)]. When the bias is increased and the hold is released, the system enters the drift phase; here, the potential barrier disappears, the mean DNA position begins to move toward the *trans* side, and the distribution broadens owing to free diffusion [Fig. 2(c)]. When the mean DNA position reaches the midpoint between two adjacent potential peaks, the hold mechanism is activated once more, initiating the second hold phase; at this stage, DNA motion is frozen again [Fig. 2(d)]. Finally, in the trap phase, the transmembrane voltage returns to the stall voltage $V_{\text{stall}}$, and the hold is released, allowing the DNA to relax toward the bottom of the potential well and recover a localized equilibrium distribution [Fig. 2(e)]. During the trap phase, in addition to relaxation, DNA may thermally escape to adjacent wells with a certain probability, which is discussed later as the trap error rate.

The essential idea here is that introducing a hold phase after the trap phase increases the effective residence time near the potential minimum, effectively eliminating the 100–1,000 μs lower limit imposed by basecalling requirements[1]. Trap times are now constrained only by the relaxation of the DNA distribution, which could reduce the limit by three to four orders of magnitude. Consequently, the required friction between the DNA and nanopore decreases within a realizable range.

Although the temporal constraint on the trap phase is alleviated, its role remains essential. While the hold phase prevents DNA leakage, it also halts the relaxation (narrowing) of the probability distribution. Without the trap phase, positional uncertainty accumulates cycle-by-cycle, eventually leading to loss of control.

We now examine the theoretical feasibility of the proposed ratchet mechanism. The analysis focuses on the drift and trap phases, estimating the corresponding durations, friction coefficients, and resulting error rates for reliable operation. We assume the DNA position to follow a Gaussian probability distribution and track its mean and variance through each phase.

For the unzipping potential, we adopt the one-nucleotide periodic potential reported by Suma et al.,[12] who used coarse-grained molecular dynamics simulations to analyze the unzipping of dsDNA at the nanopore rim and reconstructed the potential landscape under the stall voltage. To facilitate the analytical treatment, they approximated the resulting potential using a piecewise triangular function [Fig. 3]:

$$U(z) = \begin{cases} \dfrac{\Delta U}{2}\left[\cos\left(-\dfrac{\pi}{z^{\ddagger}}(z - z^{\ddagger})\right) + 1\right] & \text{for } 0 \leq z \leq z^{\ddagger} \\ \dfrac{\Delta U}{2}\left[\cos\left(\dfrac{\pi}{d - z^{\ddagger}}(z - z^{\ddagger})\right) + 1\right] & \text{for } z^{\ddagger} \leq z \leq d \end{cases}, \qquad (1)$$

where $d = 0.6$ nm represents the length of a nucleotide, $\Delta U = 18$ pN nm represents the height of the potential barrier, and $z^{\ddagger} = d/4$ represents the position of the peak. The authors further set the system temperature to $T = 300$ K, and this study uses this setting.

In the drift phase, the probability distribution of the DNA position spreads over time [Fig. 2(c)]. When the potential barrier is restored at the end of this phase, a fraction of the distribution may lie outside the potential valley ($z < d/4$ or $z > (5d)/4$); these molecules fall into adjacent wells, resulting in a drift error. Building on previous analytical treatments of drift-induced errors,[15–17] we evaluate the drift error rate.

Here, we assume that in the drift phase, instead of explicitly tilting the potential, as shown in Fig. 1(b) and 2(c), both the additional electrophoretic force $\Delta F = q_{\text{eff}}(V_{\text{dr}} - V_{\text{stall}})/d$ and the restoring force derived from the potential $F_{\text{res}}$ act on the DNA. $q_{\text{eff}}$ is the effective charge per phosphate group inside the nanopore, and we take $q_{\text{eff}} = 0.4$ e, which is typical for solid-state nanopores.[18] Given that Suma et al. reported an electrophoretic force of 17 pN at the stall state,[12] $V_{\text{stall}}$ is calculated to be 159 mV under this assumed effective charge.

We approximate the potential to simplify the drift-time calculation using a piecewise linear function in which the restoring force is constant within each segment [Fig. 3]:

$$F_{\text{res}} = \begin{cases} F_{\text{b}} = -120 \text{ pN} & \text{for } 0 \leq z \leq z^{\ddagger} \\ F_{\text{f}} = 40 \text{ pN} & \text{for } z^{\ddagger} \leq z \leq d \end{cases}. \qquad (2)$$

The mean position of the DNA reaches the midpoint $z = (3d)/4$ after a drift duration $t_{\text{dr}}$, which is expressed as a function of the drift voltage $V_{\text{dr}}$:

$$t_{\text{dr}} = \zeta d^2 \left[ \frac{1}{4(q_{\text{eff}}(V_{\text{dr}} - V_{\text{stall}}) + F_{\text{b}}d)} + \frac{2}{4(q_{\text{eff}}(V_{\text{dr}} - V_{\text{stall}}) + F_{\text{f}}d)} \right], \qquad (3)$$

where $\zeta$ denotes the friction coefficient experienced by DNA within the nanopore. The drift error rate is equal to the tail probability of a Gaussian distribution truncated at the two neighboring peaks $z = d/4, (5d)/4$:

$$P_{\text{dr,err}} = \text{erfc}\left( \frac{d/2}{\sqrt{2\sigma_{\text{eq}}^2 + 4Dt_{\text{dr}}}} \right), \qquad (4)$$

where $D$ represents diffusion coefficient of the DNA in the nanopore and $\sigma_{\text{eq}}^2$ denotes the equilibrium variance within a potential well.

In the harmonic approximation, the bottom of each potential well can be treated as a quadratic function, $U(x) \approx (1/2)U''_{\text{eff}}x^2$, where $U''_{\text{eff}}$ is the effective curvature of the potential at the valley bottom (see Supplementary Material S1). A particle confined in such a harmonic well has a Gaussian equilibrium distribution with variance $\sigma_{\text{eq}}^2 \approx (k_B T)/U''_{\text{eff}}$.

Substituting the Einstein relation $D\zeta = k_B T$ into Eqs. (3) and (4) cancels the diffusion and friction coefficients, showing that the drift error rate is independent of both of them, as also discussed by Polonsky et al.[17] Intuitively, larger friction reduces the drift velocity but simultaneously suppresses diffusion; the two effects cancel each other, leaving the drift error rate unchanged.

As shown in Fig. 4(a), the drift error rate $P_{\text{dr,err}}$ decreases exponentially with increasing drift voltage $V_{\text{dr}}$ because a higher voltage allows the DNA to reach the midpoint $z = (3d)/4$ within a shorter period, thereby suppressing the broadening of the probability distribution caused by free diffusion.

In practice, an excessive transmembrane voltage can damage the membrane or electrodes and induce electrochemical reactions such as water electrolysis. DNA translocation has been demonstrated under voltages of up to 6 V.[19] Thus, a drift voltage of 5 V is considered experimentally feasible under pulsed operation. At $V_{\text{dr}} = 5$ V, the calculated drift error rate $P_{\text{dr,err}} = 0.0058$ is smaller than the thermally activated trap error discussed later; therefore, higher voltages offer little advantage. Hence, $V_{\text{dr}} = 5$ V is adopted for the subsequent analysis.

A drift voltage of 5 V generates an electrophoretic force in the order of 500 pN. Under such a strong driving force, moving the DNA by a short distance of $(3d)/4 = 0.45$ nm requires precise timing control of the drift duration $t_{dr}$ as well as an appropriately large friction coefficient $\zeta$. Equation (3) quantifies this requirement, showing that $t_{dr}/\zeta = 9.19 \times 10^{-4}$ nm/pN [Fig. 4(b)]. The specific design values of $t_{dr}$ and $\zeta$ that meet this requirement will be discussed later.

In the above analysis, the hold mechanism is treated as an ideal reversible immobilization process capable of countering such strong electrophoretic forces. In practice, the resisting force can be increased by extending the hold electrode along the DNA contour such that the transverse electric field acts on a larger number of phosphate charges, yielding a proportional increase in the total resisting force. Further design considerations are presented in Supplementary Material S4.

In the trap phase, the probability distribution of DNA must be sufficiently relaxed such that the molecules are well localized near the bottom of the potential well [Fig. 2(e)]. Therefore, the key design consideration is the duration of the trap phase, which should be sufficiently long for the distribution to reach this equilibrium state. Once the required trap duration is determined, the probability of thermal leakage during that period, that is, the trap error rate $P_{tr,err}$, can be evaluated.

We analyze the relaxation of the probability distribution separately for its variance and mean positions. Let $\sigma_{tr}^2$ and $\sigma_{dr}^2$ denote the standard deviations at the beginning and end of the drift phase, respectively. During the drift phase, the distribution broadens due to diffusion, giving $\sigma_{dr}^2 = \sigma_{tr}^2 + 2Dt_{dr}$. In the subsequent trap phase, the distribution narrows under the confining potential, which can be described by the Ornstein–Uhlenbeck (OU) process:

$$\sigma_{tr}^2 = \sigma_{eq}^2 + (\sigma_{dr}^2 - \sigma_{eq}^2) \exp\left(-\frac{2U_{eff}'' t_{tr}}{\zeta}\right). \tag{5}$$

When the drift and trap phases alternate repeatedly, the variance follows a recurrence relationship:

$$\sigma_{tr,n+1}^2 = \sigma_{eq}^2 + (\sigma_{tr,n}^2 + 2Dt_{dr} - \sigma_{eq}^2) \exp\left(-\frac{2U_{eff}'' t_{tr}}{\zeta}\right). \tag{6}$$

The variance converges to a steady-state value:

$$\sigma_{tr,\infty}^2 = \sigma_{eq}^2 + \frac{2Dt_{dr}}{\exp\left(\frac{2U_{eff}'' t_{tr}}{\zeta}\right) - 1} = \sigma_{eq}^2 \left[1 + \frac{\frac{2U_{eff}'' t_{dr}}{\zeta}}{\exp\left(\frac{2U_{eff}'' t_{tr}}{\zeta}\right) - 1}\right]. \tag{7}$$

Next, we consider the relaxation of the mean position. During each drift phase, the mean position shifts by $-d/4$ from the potential minimum [Figs. 2(c)–(e)]. In the OU process, the mean relaxation proceeds with a time constant equal to half that governing the variance relaxation. Let the displacement of the mean DNA position from the potential minimum after the $n$-th cycle be denoted as $\bar{z}_{tr,n}$. This displacement is related to that in the previous cycle as follows:

$$\bar{z}_{tr,n+1} = \left(\bar{z}_{tr,n} - \frac{d}{4}\right) \exp\left(-\frac{U_{eff}'' t_{tr}}{\zeta}\right). \tag{8}$$

Accordingly, the mean position converges to:

$$\bar{z}_{\text{tr},\infty} = -\frac{d/4}{\exp\left(\frac{U''_{\text{eff}} t_{\text{tr}}}{\zeta}\right) - 1}. \tag{9}$$

When the exponential terms in the steady-state expressions for the variance [Eq. (7)] and mean position [Eq. (9)] are sufficiently large—specifically, when $(U''_{\text{eff}} t_{\text{tr}})/\zeta = 8$—the system can be regarded as having reached full relaxation, satisfying $\sigma^2_{\text{tr},\infty} \approx \sigma^2_{\text{eq}}$ and $\bar{z}_{\text{tr},\infty} \approx 0$. This condition is equivalent to requiring that:

$$\frac{t_{\text{tr}}}{\zeta} = \frac{8}{U''_{\text{eff}}} = \frac{4d^2}{\pi^2 \Delta U} = 8.11 \times 10^{-3} \text{ nm/pN}, \tag{10}$$

as shown in Fig. 4(b).

Next, we evaluate the trap error rate $P_{\text{tr,err}}$ associated with leakage under this condition. Let the leakage rate be $k$; then, $P_{\text{tr,err}} = 1 - \exp(-k t_{\text{tr}})$. As the potential is periodic, the forward leakage rate $k_f$ and backward leakage rate $k_b$ are equal, yielding $k = k_f + k_b = 2k_f$. For overdamped motion in water, $k_f$ is given by the Kramers reaction-rate expression in the high-friction limit:[20]

$$k_f = \frac{\sqrt{U''(z_0)|U''(z^\ddagger)|}}{2\pi\zeta} \exp\left(-\frac{\Delta U}{k_B T}\right), \tag{11}$$

where $z_0$ and $z^\ddagger$ denote the positions of the local potential minimum and maximum, respectively.

The Kramers rate expression assumes a quasi-stationary distribution within the potential well. Although the trap phase starts from a nonequilibrium state, this assumption is valid because the relaxation time is much shorter than the leakage time, as verified in Supplementary Material S2.

By substituting the effective curvature $U''_{\text{eff}} = (2\pi^2 \Delta U)/d^2$ (see Supplementary Material S1) for $U''(z_0)$ and $|U''(z^\ddagger)|$ in Eq. (11), we obtain:

$$k_f = \frac{\pi \Delta U}{\zeta d^2} \exp\left(-\frac{\Delta U}{k_B T}\right). \tag{12}$$

Therefore, the trap error rate is given by:

$$P_{\text{tr,err}} = 1 - \exp(-2k_f t_{\text{tr}}) = 1 - \exp\left(-\frac{2\pi \Delta U}{d^2} \frac{t_{\text{tr}}}{\zeta} \exp\left(-\frac{\Delta U}{k_B T}\right)\right). \tag{13}$$

Substituting $t_{\text{tr}}/\zeta$ from Eq. (10) gives:

$$P_{\text{tr,err}} = 1 - \exp\left(-\frac{8}{\pi} \exp\left(-\frac{\Delta U}{k_B T}\right)\right) = 0.032. \tag{14}$$

Combining $P_{\text{tr,err}}$ and $P_{\text{dr,err}}$ gives a total error rate of 0.038 per ratcheting cycle, which is sufficient for sequencing based on ionic current blockades. Because typical nanopores sense 3–12 nucleotides simultaneously,[1] the current signal inherently reflects multiple neighboring bases, providing redundancy that allows basecallers to correct ratcheting errors.[21] Motor proteins used in biological nanopore sequencing show the deletion or duplication errors of 1–10%,[21,22] comparable with the present method. Moreover, unlike the stochastic stepping of motor proteins, the timing of each ratchet step is deterministic, offering a valuable cue for basecalling and potentially improving sequencing accuracy. These results suggest that a solid-state nanopore sequencer equipped with the proposed ratchet mechanism can outperform other biological systems in terms of accuracy.

Theoretically, $t_{\mathrm{dr}}$, $t_{\mathrm{tr}}$, and $\zeta$ may be chosen anywhere along the two design lines in Fig. 4(b), which give identical ratcheting accuracy. In practice, however, additional implementation constraints must be considered.

The first limitation concerns the lower bounds of $t_{\mathrm{dr}}$ and $t_{\mathrm{tr}}$. Both durations are determined by the on–off switching of the hold mechanism. According to molecular dynamics simulations by Luan et al., a hold mechanism based on electrostatic adsorption can operate in on–off cycles of approximately 0.1 μs,[13] which, therefore, defines a practical lower limit for both $t_{\mathrm{dr}}$ and $t_{\mathrm{tr}}$.

The second limitation concerns the upper bound of the friction coefficient $\zeta$. Experiments have shown that friction can be enhanced by 10–30-fold by tuning the salt concentration or viscosity gradients and by up to 100-fold through structural modification of the nanopore with nanofiber meshes or hydrogels.[3] In addition, both experiments and simulations indicate that introducing a surface charge on the nanopore wall slows DNA translocation,[23–25] suggesting that the same electrostatic mechanism could enhance friction and provide the hold function [Fig. 1(a)]. Combining these approaches could raise the effective friction to approximately 100 pN μs/nm—roughly 2–20-fold that of the protein pore α-hemolysin (see Supplementary Material S3).

Considering the practical constraints on both switching time and friction discussed above, a realistic design example is indicated by the dotted line in Fig. 4(b), corresponding to $t_{\mathrm{dr}} = 0.1$ μs, $t_{\mathrm{tr}} = 1$ μs, and $\zeta = 100$ pN μs/nm. This parameter set provides a feasible balance between performance and implementation limits.

In summary, we proposed a protein-independent ratchet mechanism for solid-state nanopore sequencing. Long and controllable dwell times are achieved without the assistance of motor proteins by coupling the moderate unzipping barrier of dsDNA with a reversible hold mechanism. The theoretical analysis indicates that a hold device operating at submicrosecond switching speeds can achieve ratcheting with an error rate below 5%. This accuracy is comparable with that achieved with motor-protein-driven ratcheting, while offering deterministic timing, which is a unique advantage of basecalling. This concept opens a promising route toward solid-state nanopore sequencers.

Key next steps include identifying hold-mechanism architectures that can operate on sub-microsecond timescales while withstanding ~500 pN electrophoretic forces, and developing methods to increase nanopore–DNA friction. Alternatively, sub-microsecond transmembrane-voltage switching—building on the 6-μs pulses demonstrated in SBX[5]—may offer a viable route to achieve similar control (see Supplementary Material S5).

# SUPPLEMENTARY MATERIAL

The supplementary material provides additional theoretical details supporting the main analysis. It includes the derivation of the effective curvature used in the Ornstein–Uhlenbeck and Kramers formulations, a validation of the high-friction Kramers rate under trap-phase conditions, an estimate of the friction coefficient of α-hemolysin for reference, design considerations for achieving large resisting forces in the hold mechanism, and an analysis of the feasibility of transmembrane-voltage pulsing as an alternative means for drift-time control.

# ACKNOWLEDGMENTS

The author thanks Minoru Kashihara and Motohide Yasuno for their helpful discussions. The author also thanks Editage (www.editage.jp) for English language editing.

# CONFLICT OF INTEREST STATEMENT

The author has no conflicts to disclose.

# AUTHOR CONTRIBUTIONS

Tomoki Ohkubo – Conceptualization; formal analysis; writing – original draft preparation; writing – review and editing.

# DATA AVAILABILITY

Data sharing is not applicable to this article as no new data were created or analyzed in this study.

Manuscript+SI_DigitalUnzipping_260316_202.docx10. J. Muzard, M. Martinho, J. Mathé, U. Bockelmann, and V. Viasnoff, Biophys. J. **98**(10), 2170–2178 (2010).
11. B. Luan and R. Zhou, USA Patent No. 9518294 (2016).
12. A. Suma, V. Carnevale, and C. Micheletti, Phys. Rev. Lett. **130**(4), 048101 (2023).
13. B. Luan, C. Wang, A. Royyuru, and G. Stolovitzky, Nanotechnology **25**(26), 265101 (2014).
14. M. Shankla and A. Aksimentiev, Nat. Commun. **5**, 5171 (2014).
15. L.P. Faucheux, L.S. Bourdieu, P.D. Kaplan, and A.J. Libchaber, Phys. Rev. Lett. **74**(9), 1504–1507 (1995).
16. J.S. Bader, R.W. Hammond, S.A. Henck, M.W. Deem, G.A. McDermott, J.M. Bustillo, J.W. Simpson, G.T. Mulhern, and J.M. Rothberg, Proc. Natl. Acad. Sci. U.S.A. **96**(23), 13165–13169 (1999).
17. S. Polonsky, S. Rossnagel, and G. Stolovitzky, Appl. Phys. Lett. **91**(15), 153103 (2007).
18. V. Tabard-Cossa, M. Wiggin, D. Trivedi, N.N. Jetha, J.R. Dwyer, and A. Marziali, ACS Nano **3**(10), 3009–3014 (2009).
19. J.B. Heng, A. Aksimentiev, C. Ho, P. Marks, Y.V. Grinkova, S. Sligar, K. Schulten, and G. Timp, Nano Lett. **5**(10), 1883–1888 (2005).
20. P. Hänggi, P. Talkner, and M. Borkovec, "Reaction-rate theory: fifty years after Kramers," Rev. Mod. Phys. **62**(2), 251 (1990).
21. W. Mao, S.N. Diggavi, and S. Kannan, IEEE Trans. Inf. Theory **64**(4), 3216–3236 (2018).
22. A.H. Laszlo, I.M. Derrington, B.C. Ross, H. Brinkerhoff, A. Adey, I.C. Nova, J.M. Craig, K.W. Langford, J.M. Samson, R. Daza, K. Doering, J. Shendure, and J.H. Gundlach, Nat. Biotechnol. **32**(8), 829–833 (2014).
23. Y. He, M. Tsutsui, C. Fan, M. Taniguchi, and T. Kawai, ACS Nano **5**(7), 5509–5518 (2011).
24. M. Tsutsui, Y. He, M. Furuhashi, S. Rahong, M. Taniguchi, and T. Kawai, Sci. Rep. **2**, 394 (2012).
25. Y. Liu and L. Yobas, ACS Nano **10**(4), 3985–3994 (2016).
9

# Figures

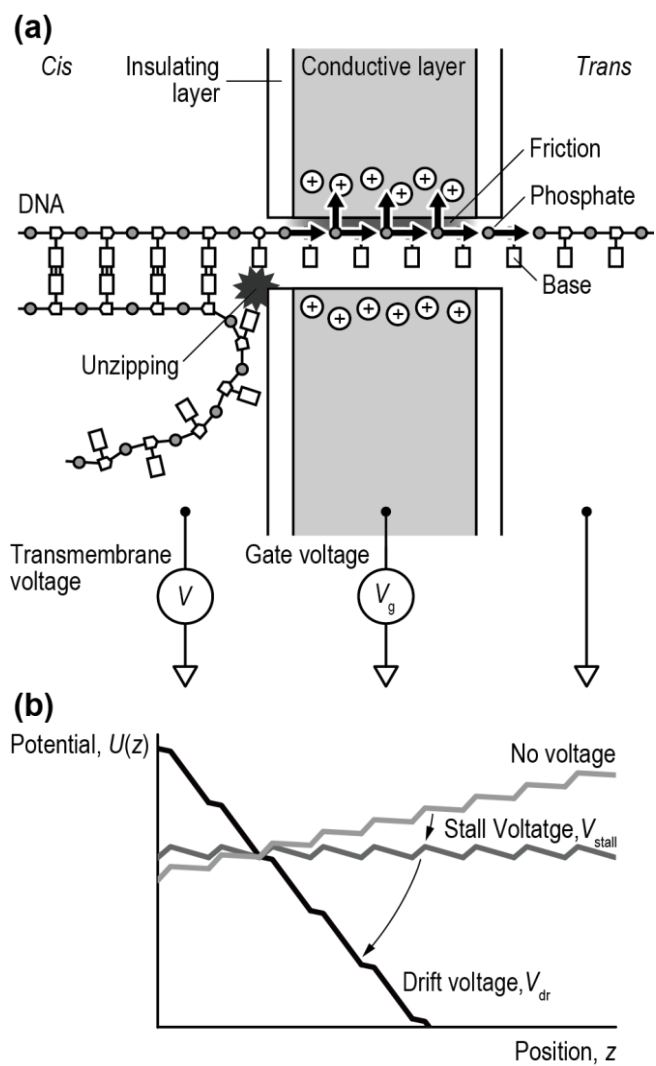

**FIG. 1. Schematic of the ratchet mechanism. (a)** Structure of the nanopore. A dsDNA molecule is electrophoretically driven toward the pore, where the duplex unzips at the rim. A conductive layer embedded in the membrane can be positively charged to attract the negatively charged DNA backbone, creating a reversible hold state. **(b)** Unzipping potential along the DNA axis. In the absence of a bias, the double-stranded state is more stable, and the potential slopes toward the *cis* side. Applying a transmembrane voltage tilts the potential toward the *trans* side; once it exceeds a threshold, the unzipping barrier disappears and translocation begins.

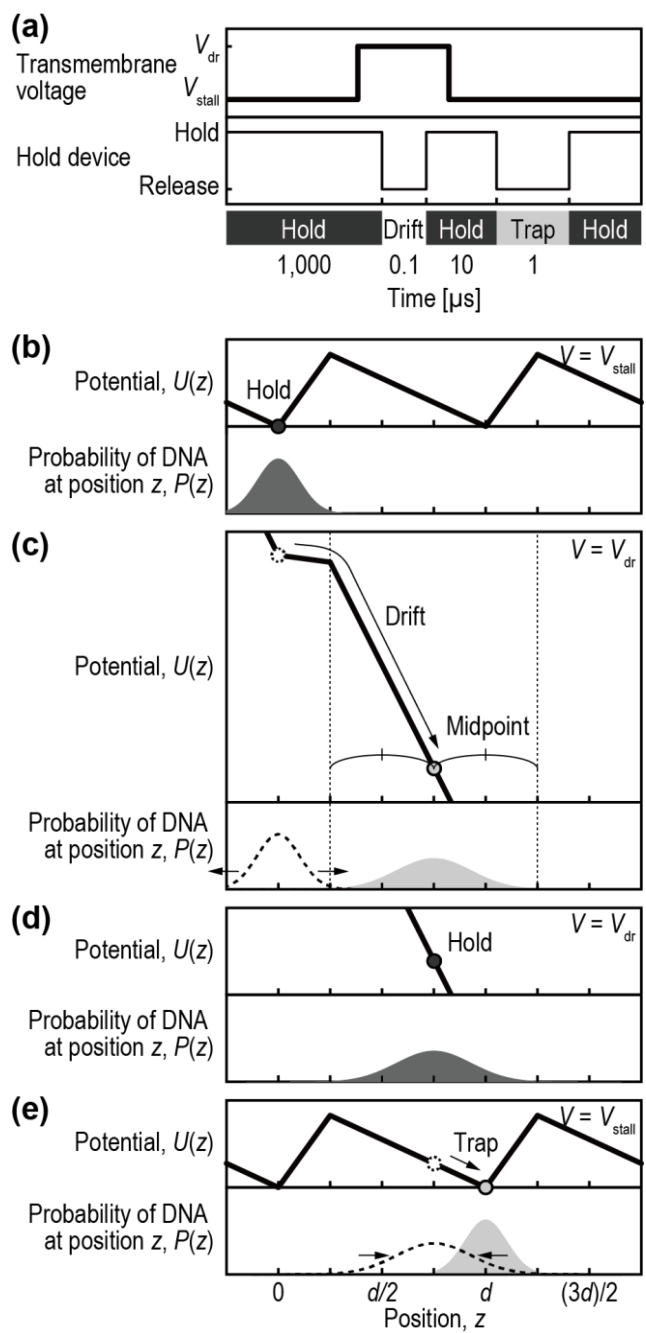

**FIG. 2. Operation sequence of the ratchet mechanism. (a)** Timing sequence of the transmembrane voltage and hold mechanism, showing four phases: the first hold, drift, second hold, and trap phases. **(b)–(e)** Potential profiles and DNA position distributions in each phase. The hold phases immobilize the DNA, the drift phase allows motion toward the *trans* side as the potential barrier disappears, and the trap phase relaxes the distribution near the potential minimum while permitting occasional thermal escape.

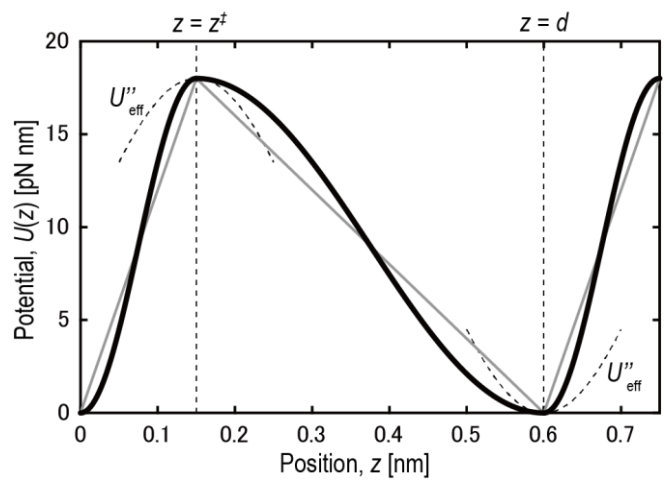

**FIG. 3. Unzipping potential function at the stall voltage**. Potential $U(z)$ for nanopore unzipping, as estimated by Suma et al. [Eq. (1)] (solid line). Because the curvature is discontinuous at the extrema, an effective curvature $U''_{\text{eff}}$ (dotted lines) is defined. For drift-time calculations, $U(z)$ is further approximated by a piecewise linear function (thin line).

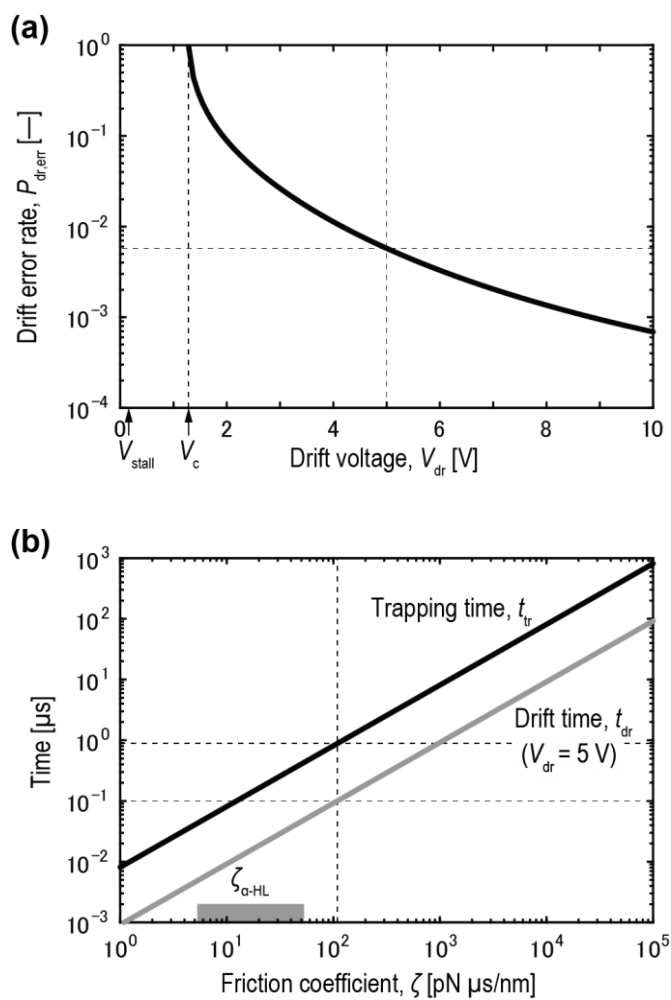

**FIG. 4. Design parameters of the ratchet mechanism. (a)** Relationship between the drift voltage $V_{\text{dr}}$ and drift error rate $P_{\text{dr,err}}$. At the critical voltage $V_c$, the potential barrier disappears and the DNA begins to drift. **(b)** Relations among the friction coefficient $\zeta$, trap time $t_{\text{tr}}$, and drift time $t_{\text{dr}}$. The gray square

marks the friction coefficient of the protein nanopore α-hemolysin (α-HL) for reference (see Supplementary Material S3), and the dotted line indicates a feasible design example corresponding to $t_{\mathrm{dr}} = 0.1$ μs, $t_{\mathrm{tr}} = 1$ μs, and $\zeta = 100$ pN μs/nm.

# SUPPLEMENTARY MATERIAL:
# Digital unzipping of DNA through a solid-state nanopore: A theoretical study for base-by-base ratcheting

Tomoki Ohkubo*
Technology Research Laboratory, Shimadzu Corporation, 3-9-4 Hikaridai, Seika-Cho, Soraku-Gun, Kyoto, 619-0237, Japan
*Correspondence: tohkubo@shimadzu.co.jp

## S1. Definition of the effective curvature

The theoretical analysis is based on the Ornstein–Uhlenbeck (OU) process and Kramers reaction-rate theory, both of which are derived from the Fokker–Planck (FP) equation. In the FP framework, the curvature at the extremes of the potential characterizes molecular motion. Therefore, we evaluate the curvature of the unzipping potential proposed by Suma et al.[1] [Eq. (1)].

Because the piecewise triangular potential function is symmetric about its peaks and valleys, only the curvature at the bottom of the valley $U''(z_0)$ must be considered. The valley corresponds to the junction of the two segments, where the curvature is discontinuous. In such cases, defining the *effective curvature* as the square of the harmonic mean of the square roots of the curvatures on both sides is appropriate:

$$U''_{\text{eff}} \coloneqq \left( \frac{2}{\frac{1}{\sqrt{U''_-}} + \frac{1}{\sqrt{U''_+}}} \right)^2. \tag{S1}$$

Substituting the curvatures obtained from the piecewise triangular potential [Eq. (1)] yields:

$$U''_{\text{eff}} = \frac{2\pi^2 \Delta U}{d^2}, \tag{S2}$$

where $\Delta U$ is the potential barrier height and $d$ is the nucleotide spacing. This definition is equivalent to approximating the valley region using a single sinusoidal function whose period is the arithmetic mean of the adjacent segments [Fig. 3, dotted lines].

The origin of the harmonic-mean form of Eq. (S1) can be derived from the Kramers reaction-rate expression in the high-friction limit. Under the assumption of a sufficiently high barrier ($\Delta U/(k_B T) \gg 1$) and quasi-stationary state, the rate constant equals the inverse of the mean first-passage time derived from the one-dimensional FP (Smoluchowski) equation:

$$t(z) = \frac{\zeta}{k_B T} \int_z^b dx \exp\left(\frac{U(x)}{k_B T}\right) \int_a^x dy \exp\left(-\frac{U(y)}{k_B T}\right), \tag{S3}$$

where $a$ and $b$ denote the reflecting and absorbing boundaries, respectively.[2] The inner integral is dominated by the vicinity of the potential minimum $z_0$, where the potential is approximated quadratically:

$$U(y) \approx U(z_0) + \frac{1}{2}U''(z_0)(y - z_0)^2. \tag{S4}$$

Because the tails of the Gaussian decay exponentially, the integration limits can be extended to $\pm\infty$. Applying the Laplace method gives:

$$\int_a^x dy \exp\left(-\frac{U(y)}{k_BT}\right) \approx \exp\left(-\frac{U(z_0)}{k_BT}\right) \int_{-\infty}^{\infty} dy \exp\left(-\frac{U''(z_0)}{2k_BT}(y - z_0)^2\right) = \exp\left(-\frac{U(z_0)}{k_BT}\right) \sqrt{\frac{2\pi k_BT}{U''(z_0)}}. \tag{S5}$$

Likewise, the outer integral dominated by the vicinity of the potential maximum $z^\ddagger$ yields:

$$\int_{z_0}^b dx \exp\left(\frac{U(x)}{k_BT}\right) \approx \exp\left(\frac{U(z^\ddagger)}{k_BT}\right) \int_{-\infty}^{\infty} dx \exp\left(-\frac{|U''(z^\ddagger)|}{2k_BT}(x - z^\ddagger)^2\right) = \exp\left(\frac{U(z^\ddagger)}{k_BT}\right) \sqrt{\frac{2\pi k_BT}{|U''(z^\ddagger)|}}. \tag{S6}$$

Combining these results leads to the Kramers rate expression in the high-friction limit:

$$k \approx \frac{1}{t(z)} = \frac{\sqrt{U''(z_0)|U''(z^\ddagger)|}}{2\pi\zeta} \exp\left(-\frac{\Delta U}{k_BT}\right). \tag{S7}$$

If the curvature near the bottom of the valley differs on the two sides, denoted by $U''_-(z_0)$ and $U''_+(z_0)$, the inner Gaussian integral in Eq. (S5) can be divided into two half-space integrals:

$$\int_a^x dy \exp\left(-\frac{U(y)}{k_BT}\right) \approx \exp\left(-\frac{U(z_0)}{k_BT}\right) \left[\int_{-\infty}^0 ds \exp\left(-\frac{U''_-(z_0)}{2k_BT}s^2\right) + \int_0^{\infty} ds \exp\left(-\frac{U''_+(z_0)}{2k_BT}s^2\right)\right]$$
$$= \exp\left(-\frac{U(z_0)}{k_BT}\right) \sqrt{2\pi k_BT} \frac{1}{2}\left(\frac{1}{\sqrt{U''_-(z_0)}} + \frac{1}{\sqrt{U''_+(z_0)}}\right), \tag{S8}$$

where $s = y - z_0$. Defining the effective curvature according to Eq. (S1) restores the same functional form as Eq. (S5):

$$\int_a^x dy \exp\left(-\frac{U(y)}{k_BT}\right) = \exp\left(-\frac{U(z_0)}{k_BT}\right) \sqrt{\frac{2\pi k_BT}{U''_{\text{eff}}(z_0)}}. \tag{S9}$$

Thus, the use of the squared harmonic mean ensures continuity between regions of differing curvatures and maintains consistency with the Kramers rate formulation. The same treatment applies to the curvature discontinuities at the top of the potential barrier.

## S2. Validity of the Kramers rate expression

The Kramers reaction-rate expression in the high-friction limit assumes that the probability distribution within a potential well is quasi-stationary. In the trap phase of the present model, the distribution initially deviates from equilibrium; therefore, whether intrawell relaxation occurs much faster than barrier crossing must be verified.

The relaxation time constant of the OU process inside the well is given by Eqs. (7) and (12): $\tau_{\text{relax}} = \zeta/U''_{\text{eff}}$, where $U''_{\text{eff}}$ is the local curvature of the potential at the maxima and minima (see Supplementary Material S1).

The characteristic time for thermally activated escape is the inverse of the Kramers rate [Eq. (15)]:

$$\tau_{\text{leak}} = \frac{1}{2k_f} = \frac{\pi\zeta}{U''_{\text{eff}}} \exp\left(\frac{\Delta U}{k_B T}\right). \tag{S10}$$

Therefore, the ratio between these two time scales is:

$$\frac{\tau_{\text{leak}}}{\tau_{\text{relax}}} = \pi \exp\left(\frac{\Delta U}{k_B T}\right) = 243. \tag{S11}$$

Hence, any nonstationary component in the intrawell distribution rapidly relaxes to a local equilibrium long before escape occurs. This evaluation justifies the use of the quasi-stationary approximation and validates the application of the high-friction Kramers rate expression in the present analysis.

## S3. Friction coefficient through α-hemolysin (α-HL)

To provide a reference for evaluating the frictional magnitude, we estimate the friction coefficient of DNA translocation through α-hemolysin (α-HL). Reported translocation velocities for α-HL at typical transmembrane voltages (∼100 mV) include 1.2 nt/μs,[3] 0.5–1 nt/μs,[4] 0.03–0.15 nt/μs,[5] and 0.12 nt/μs.[6] From these data, the average translocation velocity $v_{\alpha\text{HL}}$ can be reasonably assumed to fall within 0.1–1 nt/μs. The effective charge per nucleotide in α-HL is approximately 0.1 e/nt.[7] The electrophoretic force on DNA inside the pore is therefore expressed as

$$F = \frac{q_{\text{eff}} V}{d}, \tag{S12}$$

where $V$ is the transmembrane bias and $d$ is the nucleotide spacing (0.6 nm). Using this relation, the friction coefficient is estimated as

$$\zeta_{\alpha\text{HL}} = \frac{F}{v_{\alpha\text{HL}}} = \frac{q_{\text{eff}} V}{d v_{\alpha\text{HL}}}, \tag{S13}$$

yielding 5.3–53 pN μs/nm. This value is used as a benchmark for the analyses presented in the main text.

## S4. Design considerations for increasing the resisting force of the hold mechanism

In this study, the hold mechanism is modeled as an ideal reversible immobilization process capable of countering electrophoretic forces of several hundred piconewtons. This section presents additional design considerations that justify the physical feasibility of achieving such resisting forces without altering the effective sensing length of the nanopore.

A straightforward approach to increase the resisting force is to extend the hold electrode along the DNA contour such that the positive charge induced by the gate voltage interacts with a larger number of phosphate charges, yielding a proportional increase in total resisting force. Importantly, extending the electrode in this manner does *not* necessarily require extending the nanopore itself; the ionic current-sensing region remains as short as that in conventional solid-state nanopores.

A possible implementation is illustrated in Fig. S1. In this geometry, a gate electrode is placed in the open *cis*-side region rather than inside the pore interior. The nanopore is positioned close to the electrode such that the DNA strand naturally remains within the electric double layer of the electrode. Because the electrode resides outside the nanopore interior, the effective pore length that governs single-nucleotide discrimination in ionic-current sensing is preserved, whereas the electrode length and, thus, total resisting force can be increased.

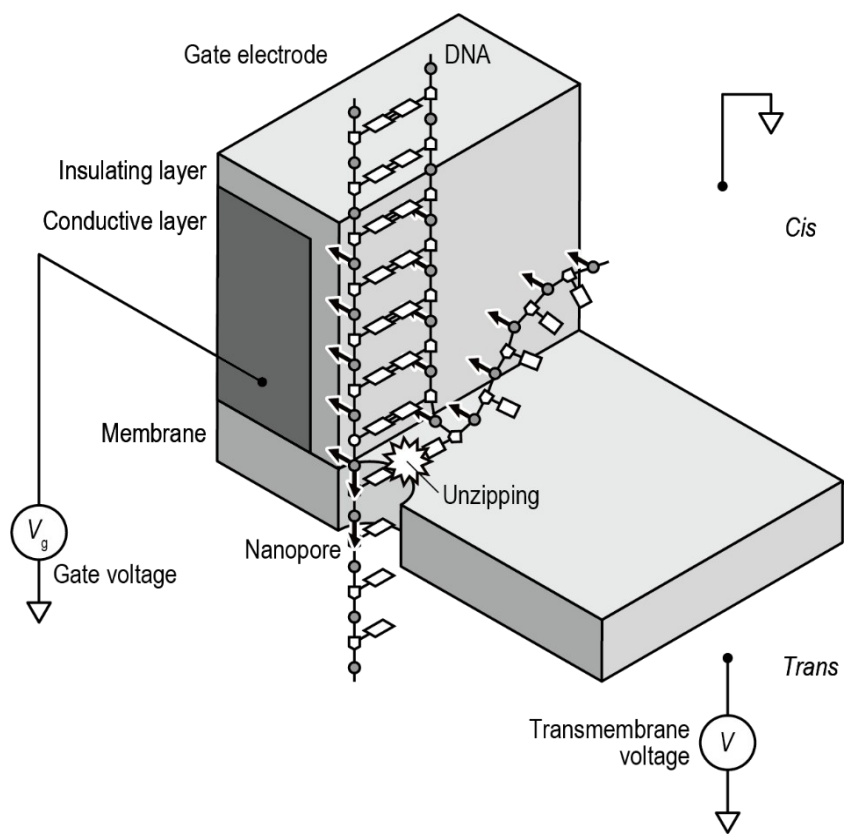

**FIG. S1. Conceptual design of a gate electrode arrangement that enhances the resisting force of the hold mechanism.** A gate electrode is placed on the *cis* side of a thin membrane. When a positive gate voltage $V_g$ is applied, the electrostatic field attracts the negatively charged phosphate backbone of DNA. The nanopore is positioned close to the electrode such that the DNA naturally remains near the charged surface. The electrode lies outside the nanopore interior, preserving the short effective pore length required for base-resolution ionic-current sensing.

## S5. Transmembrane-voltage pulsing as an alternative approach for drift-time control

In the operation sequence shown in Fig. 2 of the main text, the hold mechanism must withstand an electrophoretic force of ~500 pN immediately before and after the drift phase, when the transmembrane voltage is set to $V_{\text{dr}}$. Even with the force-amplifying strategies discussed in Supplementary Material S4, realizing a hold device capable of resisting such a load is technically challenging. A possible alternative is to control the drift duration $t_{\text{dr}}$ by rapidly switching the transmembrane voltage, rather than relying on the hold mechanism, as illustrated in Fig. S2. In this scheme, the hold mechanism no longer needs to counteract the high load.

Existing demonstrations provide a useful reference point. In sequencing-by-expansion (SBX), Kokoris et al. achieved 6-μs voltage pulses across a lipid-bilayer membrane.[8] Because solid-state membranes exhibit faster electrical response than lipid bilayers, further pulse shortening is plausible; if we assume a 1-μs pulse width for illustration, then Eq. (3) implies that the required effective friction coefficient rises to $\zeta \approx 10^3$ pN μs/nm, which corresponds to roughly 20–200 times the friction estimated for α-HL (see Supplementary Material S3). Although such high friction is demanding, it still lies within a conceivable range for solid-state nanopores employing friction-enhancing methods discussed in the main text.

These considerations suggest that high-speed transmembrane-voltage pulsing offers a plausible alternative to a strong hold mechanism, but not necessarily an easier one. Transmembrane-voltage pulsing relaxes the requirement of resisting electrophoretic forces yet imposes more stringent demands on the achievable nanopore friction. Conversely, the hold-mechanism approach alleviates friction requirements but must sustain large electrophoretic loads. Both approaches therefore represent technically challenging—but potentially feasible—routes toward achieving controlled, single-base stepping in solid-state nanopores.

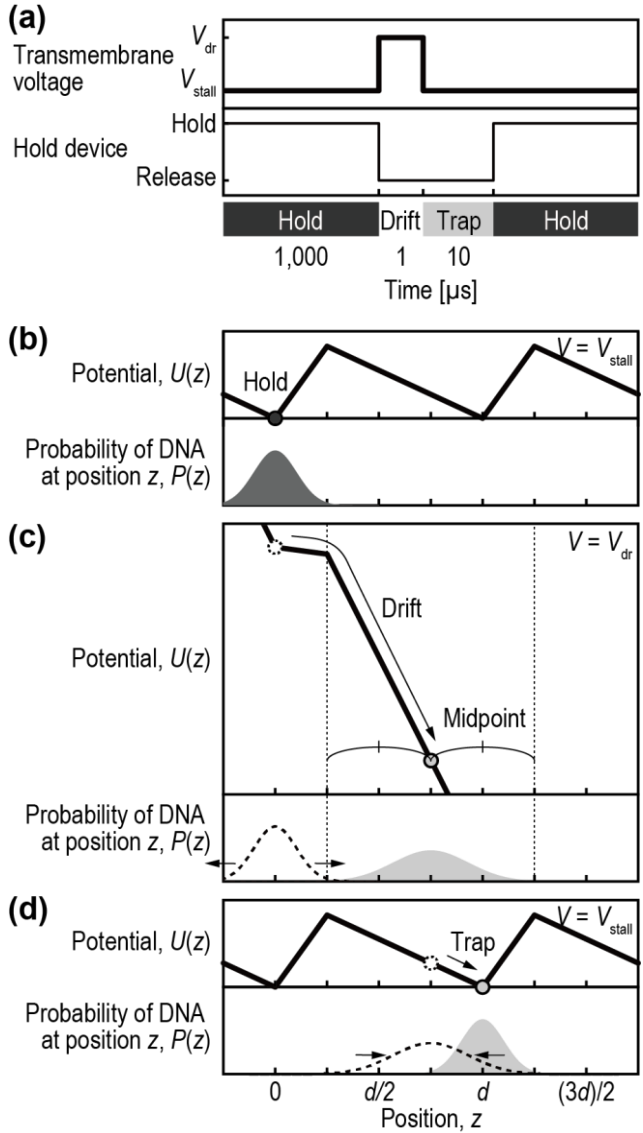

**FIG. S2. Operation of the ratchet cycle when the drift duration $t_{dr}$ is controlled by rapid switching of the transmembrane voltage.** (a) Timing diagram showing a three-phase cycle consisting of a hold phase, a drift phase, and a trap phase. In this scheme, the drift phase is initiated when the transmembrane voltage is switched from $V_{dr}$ to $V_{stall}$ simultaneously with the release of the hold mechanism, so that DNA begins drifting under the tilted potential created immediately after the voltage step. Only one hold phase is required per cycle. (b) During the hold phase, the DNA distribution remains localized near the potential minimum under the action of the hold mechanism. (c) When the voltage is stepped down and the hold mechanism is released, the potential tilts, causing the probability distribution to drift toward the midpoint over the short interval $t_{dr}$. (d) When the voltage is set to $V_{stall}$, the potential becomes periodic again and the DNA distribution relaxes into the nearest trap, completing one ratchet step. This voltage-pulse–based scheme removes the need for the hold mechanism to withstand the full electrophoretic force during the drift phase.